# LOAD FORECASTING MODEL AND DAY-AHEAD OPERATION STRATEGY FOR CITY-LOCATED EV QUICK CHARGE STATIONS

***Zeyu Liu[1], Yaxin Xie[1], Donghan Feng[1*], Yun Zhou[1], Shanshan Shi[2], Chen Fang[2]***

*[1] Key Laboratory of Control of Power Transmission and Conversion, Ministry of Education, Department of Electrical Engineering, Shanghai Jiao Tong University, 800 Dongchuan Rd., Shanghai, China*
*[2] Electric Power Research Institute, State Grid Shanghai Municipal Electric Power Company, 171 Handan Rd., Shanghai, China*
**seed@sjtu.edu.cn*

## Abstract

Charging demands of electric vehicles (EVs) are sharply increasing due to the rapid development of EVs. Hence, reliable and convenient quick charge stations are required to respond to the needs of EV drivers. Due to the uncertainty of EV charging loads, load forecasting becomes vital for the operation of quick charge stations to formulate the day-ahead plan. In this paper, based on trip chain theory and EV user behaviour, an EV charging load forecasting model is established for quick charge station operators. This model is capable of forecasting the charging demand of a city-located quick charge station during the next day, where the Monte-Carlo simulation method is applied. Furthermore, based on the forecasting model, a day-ahead profit-oriented operation strategy for such stations is derived. The simulation results support the effectiveness of this forecasting model and the operation strategy. The conclusions of this paper are as follows: 1) The charging load forecasting model ensures operators to grasp the feature of the charging load of the next day. 2) The revenue of the quick charge station can be dramatically increased by applying the proposed day-head operation strategy.

## 1 Introduction

The usage of conventional internal-combustion engine vehicles is causing environmental problems such as air pollution and global warming as well as the concern of the shortage of fossil fuel. Electric vehicles (EVs) are regarded as a solution to these problems [1]. While the charging of EV batteries has brought challenges to the operation of power system [2-3], the advantages of EVs are bringing the rapid development of EVs and the promotion of EVs by governments around the world. EVs are considered as a kind of green vehicles, which is not only because they avoid producing fossil fuel pollution, but also because of their ability to consume sustainable energies, which are environmental-friendly [4].

The usage of EVs requires the coordination of supporting facilities, among which the charging stations are the most important ones. At present, still a considerable percentage of the EV owners have no private charging posts, which indicates the tremendous need for public charging facilities as the EV market grows [5].

The load forecasting of a certain charging station is important information to the station operator. Some related studies include load forecasting for sites with general functions, but are not realistic enough for a real station [6]. A load forecasting method for quick charging stations is proposed in this paper.

For ordinary charging stations where EVs are charged in slow charging mode, the station operators are able to integrate EVs into power system by demand response. The overload of power system can be relived by discharging EVs during peak hours, where the station operators can also earn profit [7]. Other profit source for load aggregators include the day ahead scheduling in the day-ahead electricity market [8] and the ancillary services such as parking [9].

However, quick charging stations are not capable of participating in the demand response since the EVs are seeking prompt charging service here. The small-scale load of a quick charging station is not admitted in the electricity market, either. However, in some regions, the peak-valley electricity price is applied for charging facilities, which can be made use of to maximize the profit of the station [10]. Based on this, a day-ahead operation strategy for ESS (energy storage system)-equipped quick charging stations is proposed in this paper. The main contents of this paper are shown in Fig 1.

The main contributions of this paper are as follows:
(1) Offer a detailed data processing procedure for NHTS 2017 database, and extract the behaviour characteristics of household vehicles.

(2) Propose a Monte Carlo based charging load forecasting method for quick charging stations with which the charging load curve in the next 24 hours is predicted.
(3) Present an operation strategy for quick charging stations charged by valley-peak price. The ESS is involved in the optimization.

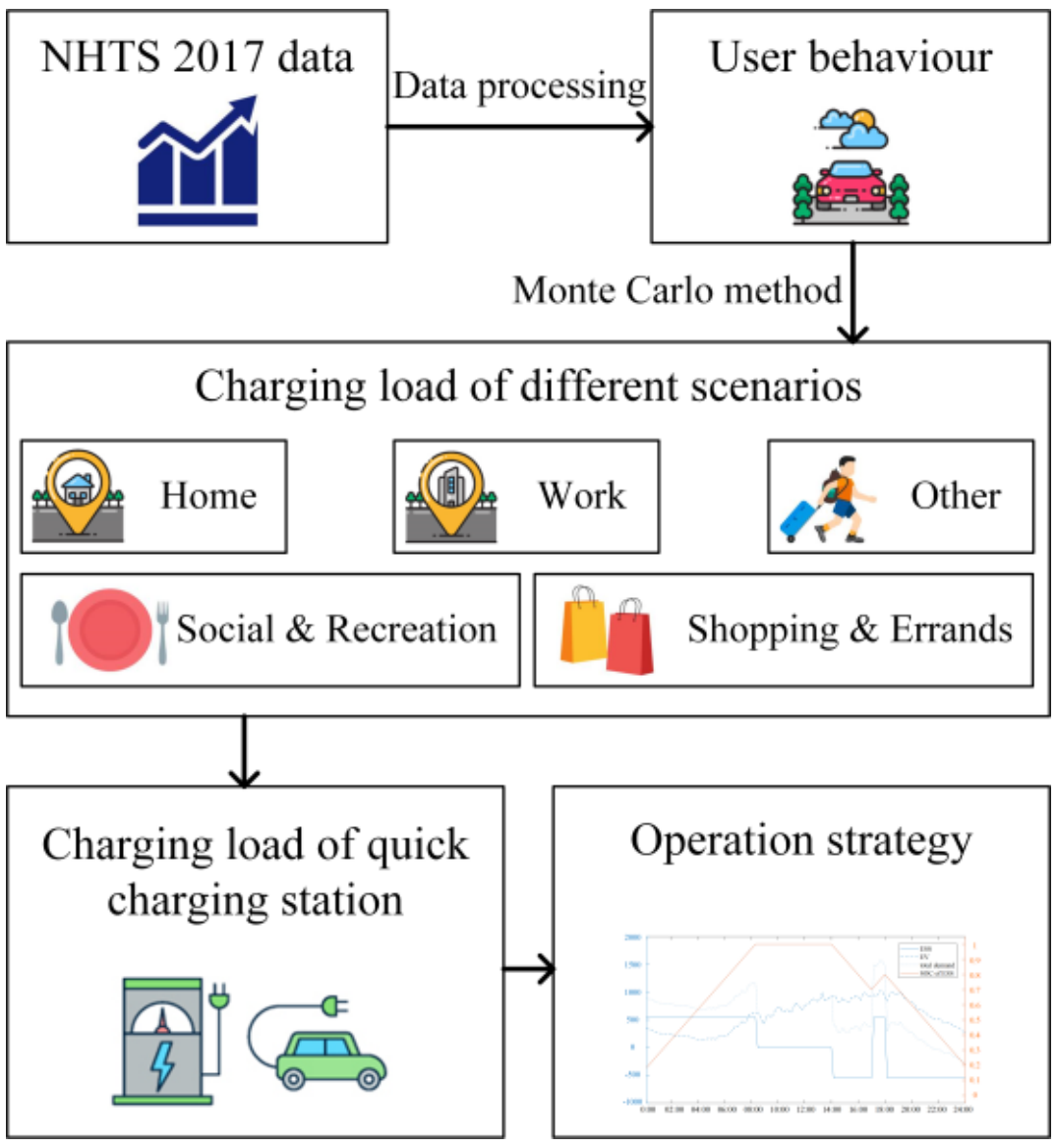

Fig 1. The main contents of this paper

## 2. Methodology

*2.1 NHTS data processing*

In the scenario of a city, the EVs can be divided into several types by their use, including household EVs, passenger EVs, freight EVs and EVs for other special purposes such as postal purpose and touristic purpose. Among these EVs, passenger EVs and freight EVs are probably charged in passenger stations or freight stations, and other EVs for special purposes are probably charged in corresponding charging facilities. However, a considerable proportion of household EV owners have no private charging post. In addition, there is a possibility that household EVs require urgent charging during the trip, since household EVs are not as scheduled as other EVs. Therefore, household EVs are the major customers of city-located quick charge stations.

Taking this fact into consideration, the charging load of quick charging stations is mainly related to household vehicles, which requires the research on household EV behaviour. Since household EVs are used for similar purposes as household petrol vehicles, it is natural to assume that the behaviour of household EVs are similar to petrol vehicles.

The destination of every household EV trip can be divided into 5 classes, which are H (home), W (work), SE (shopping & errands), SR (social & recreation), and O (other) [4]. Further, for a household EV, the owner's home is a place where the EV periodically arrives. Therefore, the trips can be artificially separated by the moment when the EV arrives home. Hence, the trips of EVs form closed chains where the owner's home is included, which is called home-based (HB) trip chains. HB trip chains can be divided into several classes by the number of trips included in the chain and the midway sites, among which trip chains with 3 or fewer trips are mainly focused on. This is because the trip chains with 4 or more trips in a day usually contain stop-by trips such as dropping children at school, where it is unlikely for the driver to get EV charged. Trip chains with 3 or fewer trips are divided into simple chains (with 2 trips) and complex chains (with 3 trips), and further divided into several types by midway sites. The classifications for trip chains are shown in Fig 2.

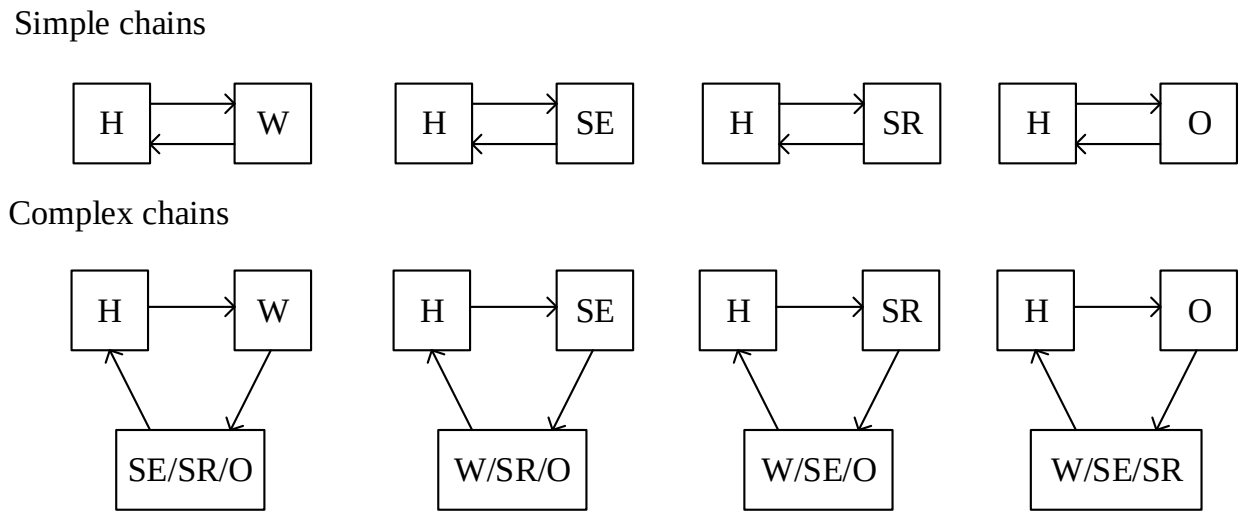

Fig. 2 Classification of trip chains

In this paper, NHTS (National Household Travel Survey) 2017 statistics are applied to analyse household EV behaviour. NHTS is conducted by the Federal Highway Administration of America, and is the authoritative source on the travel behaviour of the American public [11].

In NHTS 2017 data, the 'trippub.csv' database is used, which contains about 923500 records of household trips, where necessary information of each trip is recorded. The main information used in this research include the ID of the family and the vehicle, the trip date, the starting time, the ending time, the duration, the length, the destination and whether it is a home-based trip.

Based on NHTS 2017 statistics, the features of household EVs trips are analysed in following steps.

- Step 1. Read the data of the current travel record. For every vehicle, record the concerned data.
- Step 2. When all necessary data for a closed trip chain is recorded, add it to the data set for the corresponding trip chain type.
- Step 3. Accumulate the data for every trip of every type of trip chains in the data sets, and fit them with kernel density estimation method.

From this data processing procedure, following information is obtained:
(1) The distribution of
  a) starting/ending time
  b) trip duration
  c) trip length
  d) average velocity
in every trip of every type of trip chains.
(2) The distribution of staying duration at every midway site of every type of trip chains.
(3) The proportion of every type of trip chains.

*2.2 Load forecasting model*

Based on the information obtained by data processing, a Monte-Carlo-based load forecasting method is established. The whole trip chains of EVs are simulated in a way that is designed to be close to the real using scenarios. In the simulation, necessary parameters are generated according to the fitted distribution.

*a) Ending time of trip 1:* considering that the start time of a day's schedule is relatively fixed (e.g. time for work, appointments with clients or friends), the first moment to be generated is set to be the ending time of trip 1.

*b) Length & average velocity:* for every trip, the length is determined by the distance between the starting site and the destination, and is the major influence factor on electricity consumption, while the trip duration is affected by several factors. The average velocity, which is affected by factors including weather and traffic congestion, is the determinant for trip duration when length is fixed.

*c) Staying duration at midway sites:* the duration at the midway site is another important data in user behavior (e.g. work hours in an H-W-H trip chain). The above parameters are obtained by random extraction according to the fitting distribution of corresponding data set.

*d) Charging demand:* the principle of charging is to ensure the usage of EVs. Therefore, the EVs are set to be charged when the SOC (state of charge) is inadequate for the next trip. Considering a security redundancy of 30% SOC remained after the next trip, the charging condition is expressed as:

$$SOC_n - ul_n / p \le 0.3 \quad (1)$$

The EV's SOC at the *n*th site can be calculated by

$$SOC_n = SOC_{n-1} - ul_{n-1} \quad (2)$$

where the initial SOC ( $SOC_0$ ) is set to be 1 if the user owns a private charging post, or a random value in [0.5, 1] if the user does not.

When an EV is charged at a site, the charging power of it is added to the charge load of the quick charge station at this site during the whole charging duration. The charging duration at site $n$ is the minimum between staying time and the necessary charging time till the EV is fully charged.

$$T_{charge,n} = \min\left\{T_{stay}, (1-SOC_n)\cdot c / p\right\} \quad (3)$$

The simulation procedure can be expressed as the flow chart in Fig. 3.

Though the charging load features or all 5 kinds of sites are extracted by the above procedure, in reality, a location in a city cannot be simply classified as a single-functioned site such as a home site or a work site. The common situation is that a quick charging station is surrounded by residential buildings, office buildings, malls, hospitals, which indicates that certain proportion of functions of the city are accumulated in the effect range of the station. The final load of the station is

$$p_{station} = \sum_{i=1}^{5} p_i \cdot r_i \quad (4)$$

where $p_i$ and $r_i$ are the load and the proportion of 5 site properties in the studied area. For instance, if the proportion of H site is $p_1 = 0.04$, 4% of the city residents live in the affective area of the station.

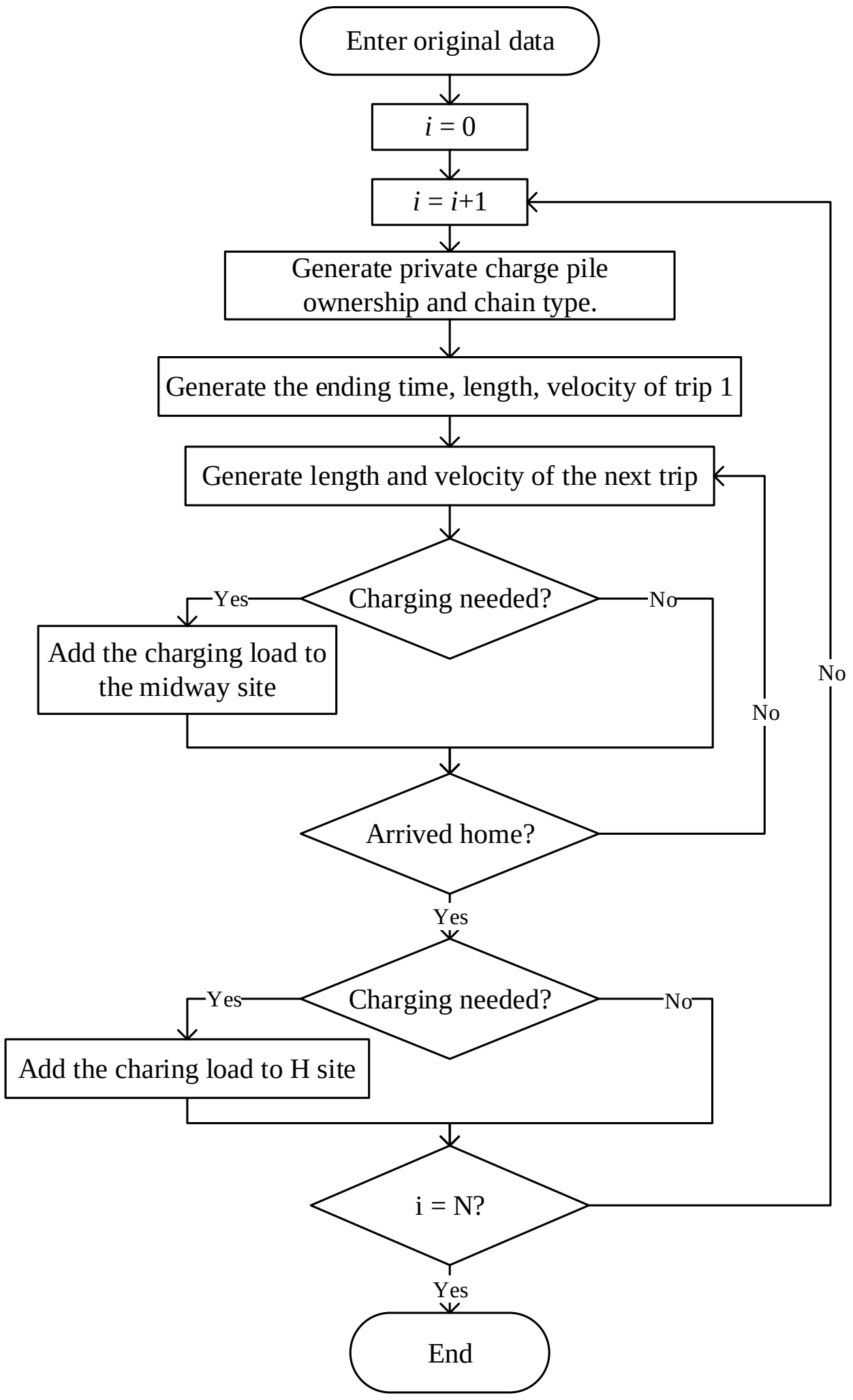

Fig 3 Flow chart for Monte Carlo simulation procedure

*2.3 Operation strategy*

In many countries and regions, the time-of-use electricity tariff is applied for EV charging stations, which means that the electricity price for station operators is higher during the peak hours. Since the peak of charging demands tend to be reached during the peak electricity price hours, a quick charging station will probably have to pay a higher electricity bill without proper operation strategy. To solve this problem, an operation strategy for ESS-equipped quick charging stations is proposed.

During the valley hours, the electricity price is lower than that in peak hours. Therefore, the total electricity cost for charging station operators will be lower if the ESS is charged during valley hours and discharged to contribute to the charging of EVs. This requires the operator to schedule the charging and discharging of the ESS. Assume that the electricity price in a day is $m(t)$, the capacity of the ESS is $c_{ESS}$, the maximal charging and discharging power of the ESS are $p_{ESS,c}$ and $p_{ESS,d}$, the charging power of EVs is $p_{EV}$, the charging power of ESS is $p_{ESS}(t)$, which is minus when the ESS is discharging. Hence, the total load of the station is

$$p_{ch}(t) = p_{ESS}(t) + p_{EV}(t) \tag{5}$$

The object function is to minimize the electricity bill during the whole 24 hours in a day. Constraints include (1) power limits: the charging and discharging power of ESS does not exceed the limits; (2) capacity limit: the SOC of ESS does not exceed 1. Therefore, the proposed optimization model can be expressed as

$$\text{min} \quad \sum p_{ch}(t) \cdot m(t) \tag{6}$$

$$\text{s.t.} \quad -p_{ESS,d} \le p_{ESS}(t) \le p_{ESS,c} \tag{7}$$

$$0 \le SOC_{ESS}(t) \le 1 \tag{8}$$

$$c_{ESS} \cdot SOC_{ESS}(t) = \sum_{k=1}^{i} p_{ESS}(t) \tag{9}$$

# 3 Case study

## *3.1 NHTS 2017 data processing results*

According to the data processing procedure, the trip features of household vehicles are extracted. Several features are observed in the results.

In Fig 4, the histogram of the ending time of trip 1 in the H-W-H trip chains is drawn in blue, and the CDF (probability density function) of the fitting distribution is drawn in red. It can be seen from the histogram that it is difficult to tell which typical distribution the ending time of tip 1 obeys. To handle this problem, kernel estimation method is applied in the data fitting. Kernel density estimation is a non-parametric estimation method which is capable of fitting the data completely based on the properties of data itself, instead of fitting with a default distribution type, which is the reason this method is applied in data fitting. The results prove that kernel density estimation has a remarkable effect in the data fitting.

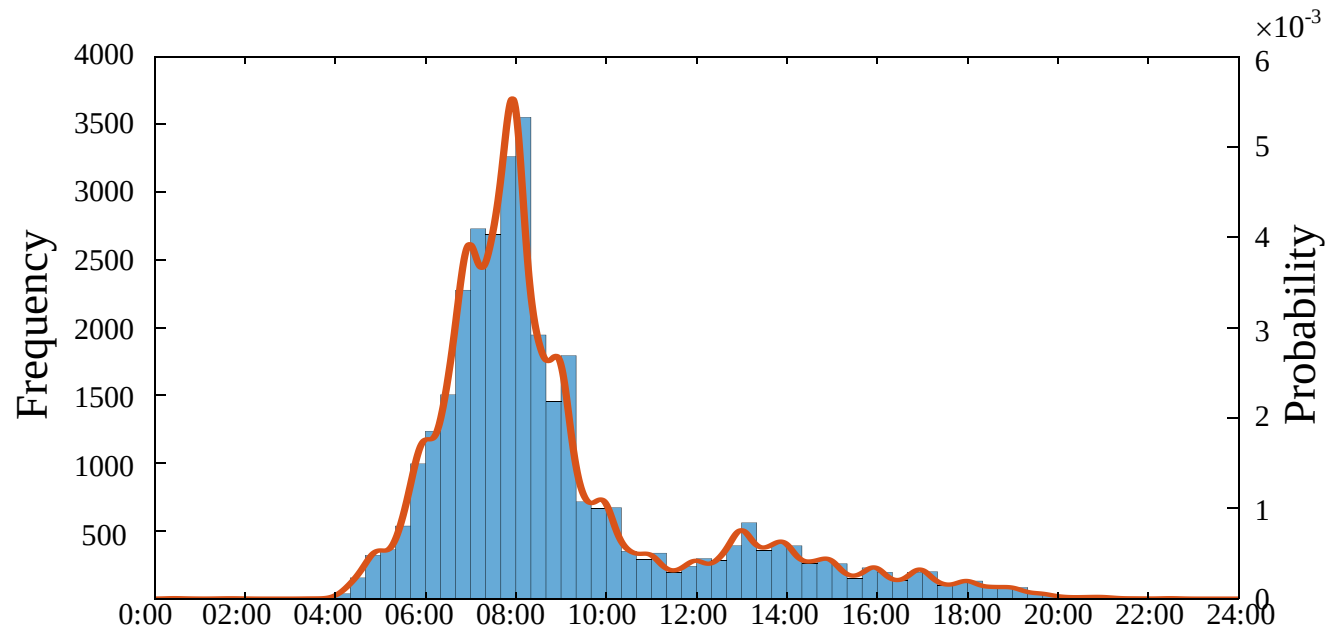

Fig. 4 Ending time of trip 1 in the H-W-H trip chains

The distributions of the ending time of trip 1 and midway site staying duration in simple trip chains are illustrated in Fig 5.

According to the distributions shown in Fig 5 (a), the probability of an EV being charged at the W site in H-W-H trip chains is obviously higher in 06:00-09:00, while the peak of H-SE-H, H-SR-H, H-O-H are reached at around 11:00, 18:00, 08:00, which is basically consistent with the life experience.

The distribution of staying duration at midway site is shown in Fig 5 (b). The staying duration at W site in H-W-H chains are most probably between 8 to 10 hours, while the midway duration of other four chains are relatively shorter, which is usually no more than 4 hours.

It is observed that the ending time of every trip and the staying duration are obviously related to trip chain types, which illustrates the rationality and the necessity of introducing different trip chains.

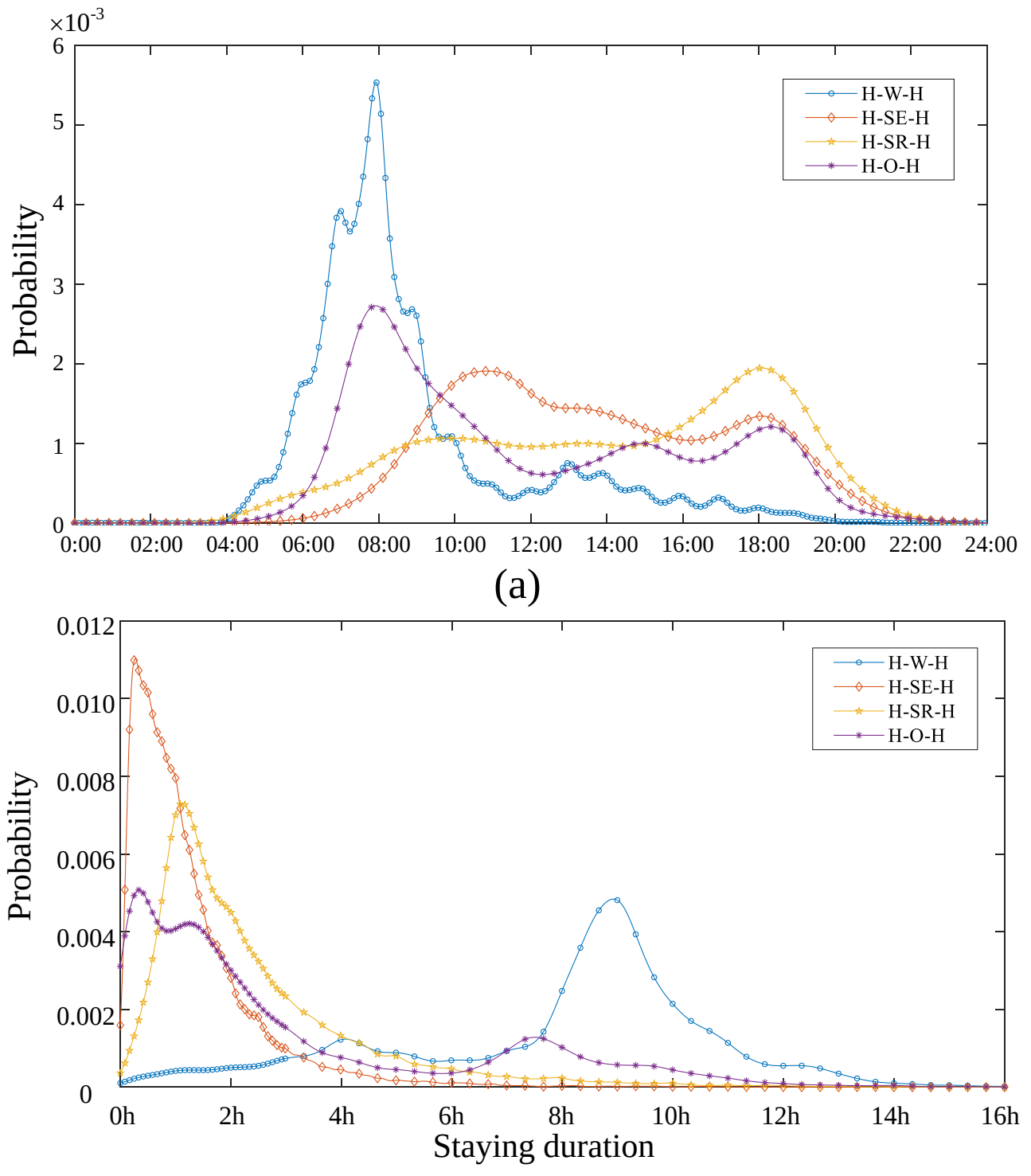

Fig 5. (a) The distribution of the ending time of trip 1, (b) the distribution of midway site staying duration in simple trip chains

## *3.2 Load forecasting*

In the Monte Carlo simulation, a 48-hour duration is considered. Since the first 24 hours are under the effect of initial conditions, only the results during the last 24 hours are taken.

The input information includes the percentage of EV uses owning private charging posts $p_{own}$, the EV number in the studied area $n_{EV}$, the charging power $p_{charing}$, the capacity of EVs $c_{EV}$, the electricity consumption per kilometre $u$ and the proportion of the 5 site properties of the station in the studied area $\boldsymbol{q}_{pro}$. The simulation input is shown in Table 1.

Table 1 Inputs of the forecasting model

| Parameter | Value |
|---|---|
| $p_{own}$ | 50% |
| $n_{EV}$ | 10000 |
| $p_{charing}$ | 60 kW |
| $c_{EV}$ | 40 kWh |
| $u$ | 0.2 kWh/km |
| $\boldsymbol{q}_{pro}$ | [0.04 0.1 0.2 0.1 0.1] |

The charging load of the 5 sites are shown in Fig 6. And the forecasting result of the charging load of the quick charging station is shown in Fig 7.

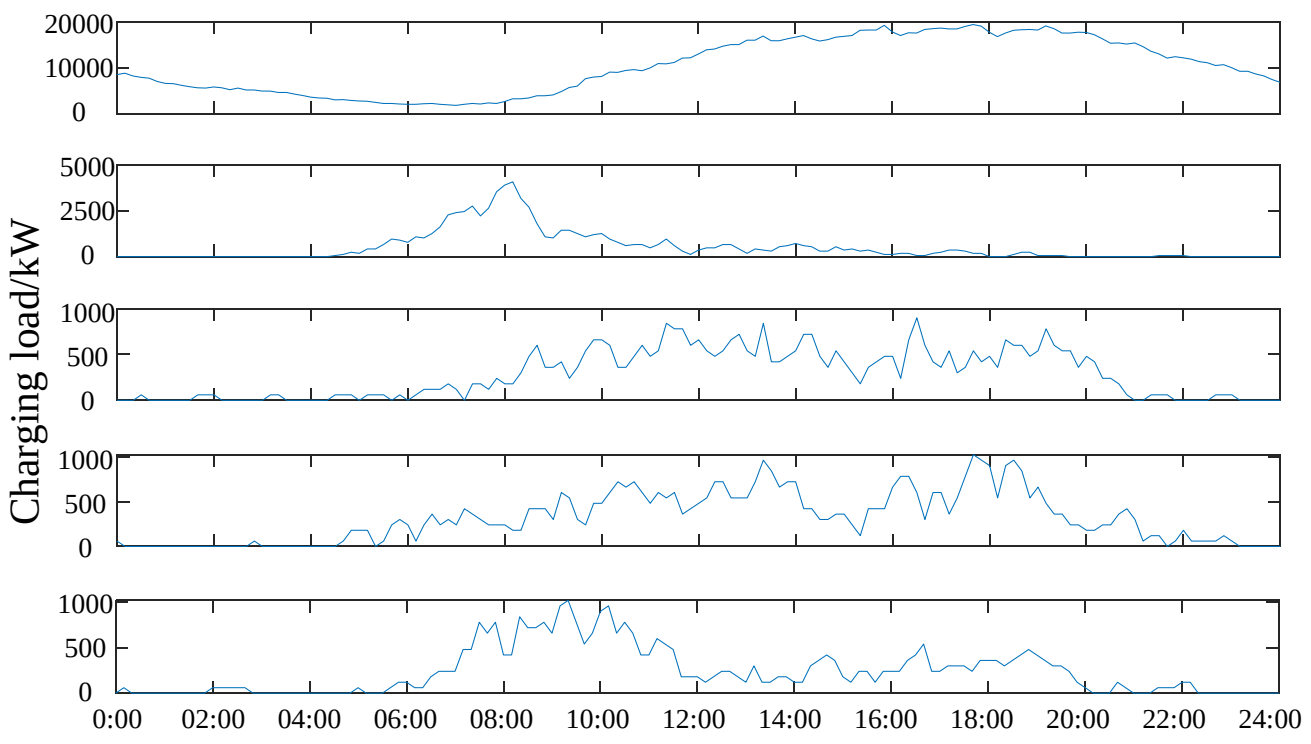

Fig 6. Charging load during a day at 5 sites

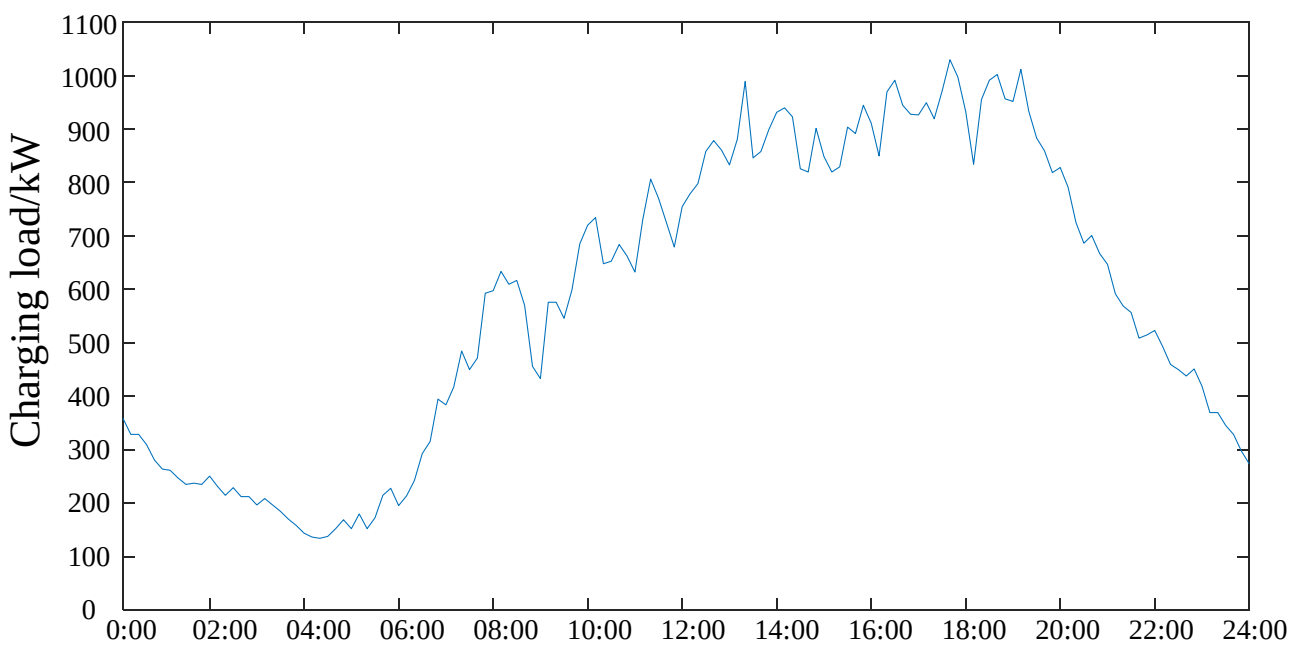

Fig 7. The charging load of the station

In Fig 6., the 5 curves are the charging load of H, W, SE, SR, and O sites in a day respectively. The existence of huge difference among the load curves at 5 sites can be observed. The different charging load have different time features, for instance, the home site had the peak charging load at around 18:00, which is the usual time when EV users finish a day's schedule and drive back home, while the peak charging demand at the work site is reached at around 08:00, which is the usual start of work hours. Besides, it is observed that the charging load at H site is much larger than that of other 4 sites, which indicates that the charging stations located in residential areas tend to have larger charging demands.

Based on the forecasting results of 5 sites presented in Fig 6, the forecasted charging load of the fictitious charging station shown in Fig 7 is acquired by (4). The operation strategy for the next day is solved based on the load forecasting.

*3.3 Operation strategy results*

In this section, the effectiveness of the operation strategy is supported by the simulation results. The parameters of this station are listed in Table 2. The price information is listed in Table 3, which is taken from Guangzhou, Guangdong Province, China [12].

Fig 8 shows the optimal strategy in the second day for a 3-day interval operation of the station. It can be seen that the ESS is charged in 00:00-08:00 and 17:00-18:00, and discharged in 14:00-17:00 and 18:00-24:00.

Table 2 Station parameters for case study

| Parameter | Value |
|---|---|
| $c_{ESS}$ /kWh | 5445 |
| $p_{ESS,c}$ /kW | 545 |
| $p_{ESS,d}$ /kW | 545 |

Table 3 Electricity price in a day

| Hours | Price/(¥/kWh) |
|---|---|
| 00:00-08:00 | 0.3338 |
| 08:00-14:00 | 0.6380 |
| 14:00-17:00 | 1.0282 |
| 17:00-19:00 | 0.6380 |
| 19:00-22:00 | 1.0282 |
| 22:00-24:00 | 0.6380 |

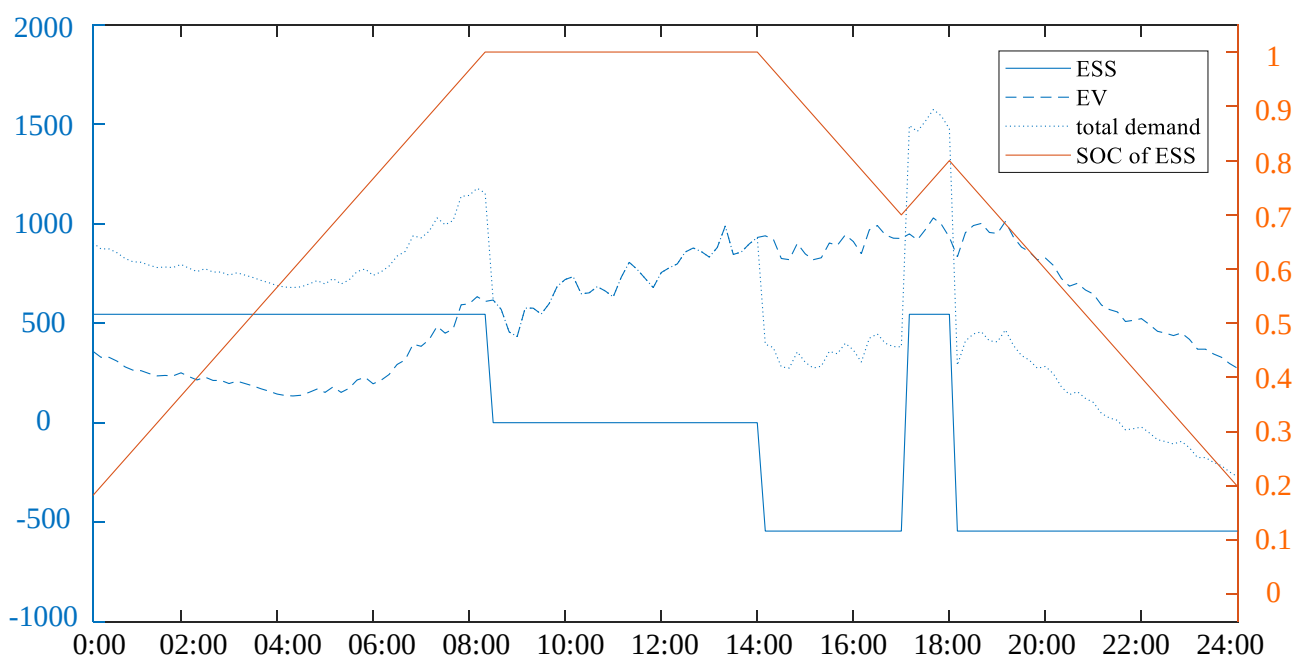

Fig 8. Operation of ESS-equipped quick charging station

According to the simulation results, generally, the ESS is charged when the electricity price is low and discharged when the price is relatively high. However, it is noticed that the ESS is not simply charged during valley hours and discharged during peak hours, which seems to be the most economic strategy for a single day. The charging scheduling of the ESS is a long-term process where the compromise of the short-term profit is needed. According to the profit of the fictious station during the 3-day interval, which is shown in Table 4, in the 3-day interval, a 24.51% reduction in electricity cost is realized by applying the proposed strategy.

Table 4 Profit of the 3-day interval in different situations

| Without ESS/¥ | Scheduled operation/¥ |
|---|---|
| 60855 | 45937 |

## 4 Conclusion

Based on the trip chain theory, an EV charging load forecasting model is established for quick charging stations, where the NHTS 2017 data is applied in the analysis of user behaviour. A detailed data processing procedure for NHTS 2017 is proposed in this paper, which gives the method of user behaviour analysis based on this database. The analysis result reveals the features of user behaviour. The Monte-Carlo-based load forecasting method is capable of forecasting the charging load of a quick charging station during a whole 24 hours, which is supported by simulation results.

Furthermore, an operation strategy for quick charging stations considering the usage of ESS and peak-valley price is proposed in this paper. The simulation result shows that, the reasonable application of ESS provides a considerable reduction on station operator’s electricity cost.

The most important contribution of this paper is that, it provides the load forecasting method and the operation strategy for quick charging stations, which is rarely mentioned in existing researches. Further researches may involve more targeted data of electric vehicles, which provides more detailed information about the EV trips. For instance, the influence factors of electricity consumption and a more refined modelling on the location of the station can be the next research direction.

## 5 Acknowledgements

The authors gratefully acknowledge the support from the Shanghai Sailing Program under Grant 19YF1423800, the Technology Program of SGCC under Grant 52094018002P, and the Participation in Research Program (PRP) of Shanghai Jiao Tong University under Grant T030PRP35071.